\documentclass[fleqn,10pt]{wlscirep}

\title{Common solar wind drivers behind magnetic storm--magnetospheric substorm dependency}

\author[1,4,5*]{Jakob Runge}
\author[2]{Georgios Balasis}
\author[2,3]{Ioannis A. Daglis}
\author[2]{Constantinos Papadimitriou}
\author[4]{Reik V. Donner}

\affil[1]{German Aerospace Center, Institute of Data Science, 07745 Jena, Germany}
\affil[2]{National Observatory of Athens, Institute for Astronomy, Astrophysics, Space Applications and Remote Sensing, Penteli, 15236 Athens, Greece}
\affil[3]{National and Kapodistrian University of Athens, Department of Physics, 15784 Athens, Greece}
\affil[4]{Potsdam Institute for Climate Impact Research, 14473 Potsdam, Germany}
\affil[5]{Imperial College, Grantham Institute, London SW7 2AZ, United Kingdom}

\affil[*]{jakobrunge@posteo.de}

\begin{abstract}
The dynamical relationship between magnetic storms and magnetospheric substorms presents one of the most controversial problems of contemporary geospace research. Here, we tackle this issue by applying a causal inference approach to two corresponding indices in conjunction with several relevant solar wind variables. We demonstrate that the vertical component of the interplanetary magnetic field is the strongest and common driver of both, storms and substorms, and explains their the previously reported association. These results hold during both solar maximum and minimum phases and suggest that, at least based on the analyzed indices, there is no statistical evidence for a direct or indirect dependency between substorms and storms. A physical mechanism by which substorms drive storms or vice versa is, therefore, unlikely.
\end{abstract}
\begin{document}

\flushbottom
\maketitle

\thispagestyle{empty}

% \section*{Introduction}

The identification of spurious associations and potentially causal relationships is key to an improved process-based understanding of various geoscientific processes.
Specifically in magnetospheric physics, the understanding of the relationship between magnetic storms and magnetospheric substorms as a part of the solar wind--magnetosphere system is of paramount importance for the development of numerical simulation models of the magnetosphere \cite{Pulkkinen2007}.
In particular, the existence and directionality of the storm -- substorm interaction is one of the most controversial aspects of magnetospheric dynamics \cite{Sharma2003}. 
The original concept of storms being the cumulative result of successive substorms put forward by Akasofu in 1961  \cite{Akasofu1961} has been disputed in subsequent analyses \cite{Gonzalez1994,Kamide1998,Sharma2003}. 
While several model-based studies have shown a distinct impact of substorm injections on the storm-time ring current enhancement \cite{Fok1999,Daglis2004,Ganushkina2005}, other studies have suggested that the ring current buildup could in principle be directly driven by the solar wind electric field \cite{Kamide1992,McPherron1988}. In this case, magnetospheric substorms do not drive magnetic storms and the two phenomena are independent and share a common cause--the southward interplanetary magnetic field (IMF) driver \cite{Daglis2003}. 

Recent work (see for instance the review by Balasis \emph{et al.}\cite{Balasis2013}) points to a considerable importance of entropy-based measures for identifying and quantifying linear and nonlinear interdependencies between different geophysical variables, variability at different scales, and other characteristics. 
Time series analyses based on information-theoretic measures have been used to shed light on the storm-substorm interaction \cite{Michelis2011} and the solar wind drivers of the outer radiation belt \cite{Wing2016} through the general perspective of quantifying information transfer, including linear and nonlinear mechanisms. In particular, DeMichelis \emph{et al.} \cite{Michelis2011} applied a bivariate transfer entropy \cite{Schreiber2000b} (bivTE) analysis to the geomagnetic activity indices AL and SYM-H. SYM-H is the high-resolution (1-min) version of the hourly Disturbance storm-time (Dst) index, which is used as a proxy of magnetospheric ring current strength and, thus, as a measure of magnetic storm intensity. AL belongs to the set of the 1-min Auroral Electrojet indices (AE, AL, AU and AO) and is used to determine the onset of the substorm growth phase \cite{Gjerloev2004}. DeMichelis \emph{et al.}  suggested that information flow from AL to SYM-H dominates in the case of small geomagnetic disturbances, while the reverse situation is observed in presence of strong geomagnetic disturbances.

However, bivariate measures such as mutual information (MI) or bivTE do not allow to exclude the very frequent influence of other variables which can act as common drivers rendering MI and bivTE associations spurious. Multivariate extensions of TE, on the other hand, are severely limited because their estimators don't work well in high dimensions \cite{Runge2012prl}. 
In the present study, we contrast bivariate measures with a directional, multivariate information-theoretic causality measure based on low-dimensionally estimated graphical models \cite{Runge2012prl,Runge2012b}. This multivariate measure for the influence of a subprocess $X$ of a system on another subprocess $Y$ is called \textit{information transfer to Y} (ITY) and allows for more powerful tests on the absence or potential presence of a causal relationship, which is crucial for developing a better ``mechanistic'' understanding of the governing processes.

Here, we investigate time series of various solar wind parameters including the IMF's magnitude $B$ and vertical component $B_Z$, its velocity $V_{SW}$ and dynamic pressure $P_{dyn}$ as well as the AL and SYM-H indices. We consider data from 2001 and 2008, near a solar activity maximum and minimum, respectively. Our goal is to clarify whether substorm activity could causally drive storm dynamics or – on the contrary – whether solar wind variables can explain the statistical associations between storm and substorm activity.

%
%%
%% Results
%%
\section*{Results}

\begin{figure*}
\centering
\includegraphics[width=.9\linewidth]{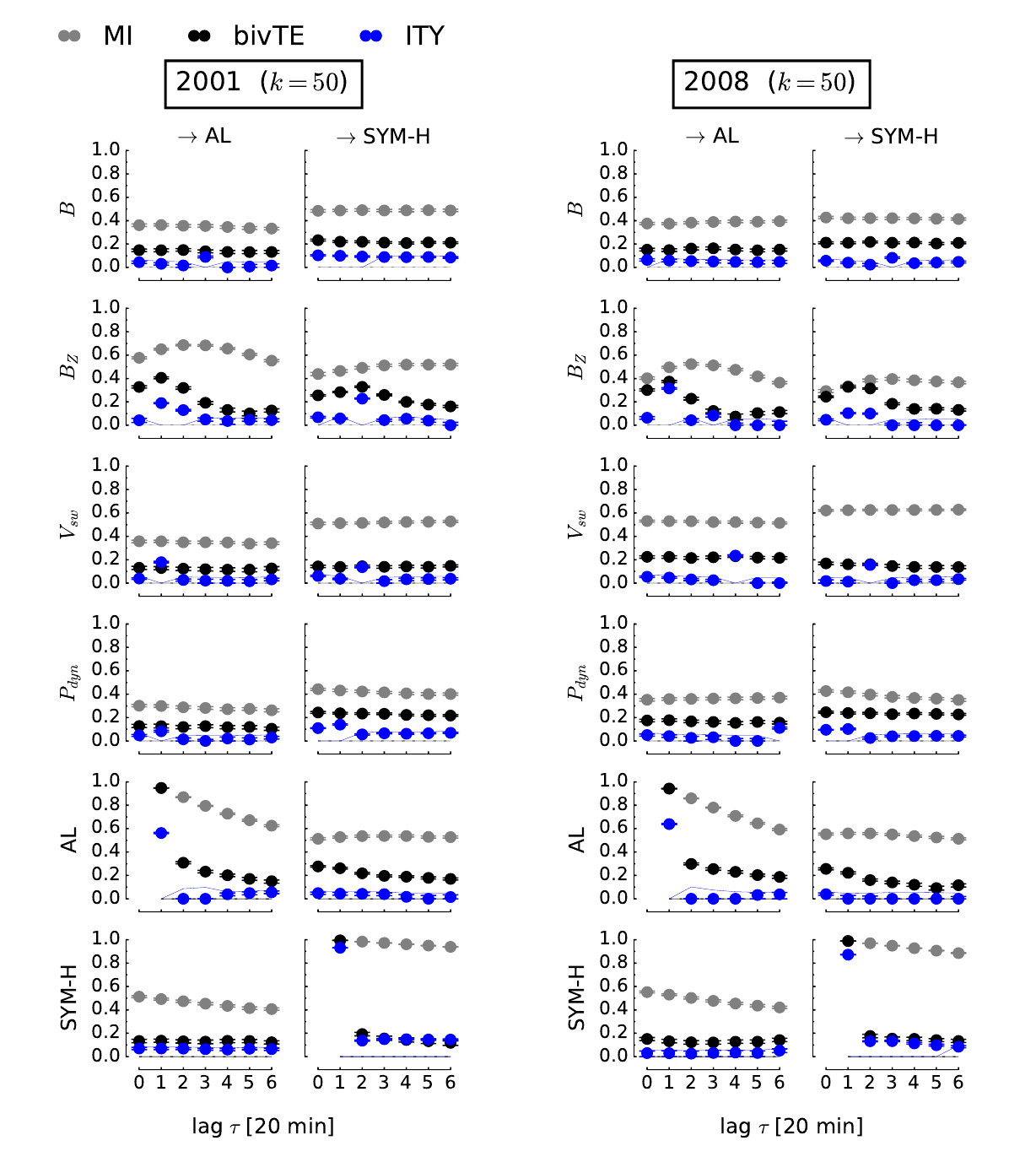}  
\caption{Lag functions of information-transfer measures (see Methods) for a year with strong (left) and weak solar activity (right). The lag functions were estimated with nearest-neighbor CMI estimation parameter $k=50$ \cite{Kraskov2004a,FrenzelPompe2007}. For example, the panel $B_Z \to AL$ shows the lag function $I(B_{Z,t-\tau}; AL_t|\ldots)$ of MI (Eq.~(\ref{eq:lagged_mi}), gray), bivTE excluding the past lag of AL (Eq.~(\ref{eq:def_te}), black), and the multivariate ITY (Eq.~(\ref{eq:def_ity}), blue) conditioning out the influence also of other variables with the parents $\mathcal{P}$ given in Tab.~\ref{tab:significance_analysis}. All (C)MI values have been rescaled to the (partial) correlation scale via $I \to\sqrt{1-e^{-2I}}\in [0,1]$ \cite{Cover2006}. For ITY, the solid line marks whether values are significant (line below dots) or not (line above dots). Significance with a computationally expensive block shuffle test was only assessed for ITY. MI and bivTE are clearly significant for a large range of lags. Confidence intervals (errorbars, mostly smaller than the dots) were estimated using bootstrap resampling involving only estimated nearest-neighbor statistics with $200$ samples.
MI and bivTE with their broad peaks clearly provide no precise information about relevant drivers and coupling delays. On the other hand, ITY features large values only at few selected lags. 
}
\label{fig:multi_causal}
\end{figure*}

% \newpage
\begin{table}
\centering
\begin{tabular}{l|c|c|c}  
\hline
\hline
\multicolumn{4}{c}{Parents of SYM-H in 2001 ($k=50$)} \\ Parent ($\tau$ [20 min.]) & $I_{\rm Par \to SYM-H}$ & $I_{\rm AL \to SYM-H}$ & $p$-value  \\ \hline
 No conds.  &  & 0.1634  & $< 10^{-2}$   \\ SYM-H (-1)  & 2.2092   & 0.0353  & $< 10^{-2}$   \\ + $B_Z$ (-2)  & 0.0572   & 0.0089  & $< 10^{-2}$   \\ + $P_{dyn}$ (-1)  & 0.0164   & 0.0022  & $< 10^{-2}$   \\ + $V_{sw}$ (-2)  & 0.0094   & 0.0010  & 0.635   \\
\hline
\multicolumn{4}{c}{Parents of SYM-H in 2008 ($k=50$)} \\ Parent ($\tau$ [20 min.]) & $I_{\rm Par \to SYM-H}$ & $I_{\rm AL \to SYM-H}$ & $p$-value  \\ \hline
 No conds.  &  & 0.1876  & $< 10^{-2}$   \\ SYM-H (-1)  & 1.9099   & 0.0253  & $< 10^{-2}$   \\ + $B_Z$ (-1)  & 0.0569   & 0.0126  & $< 10^{-2}$   \\ + $P_{dyn}$ (-1)  & 0.0161   & 0.0041  & $< 10^{-2}$   \\ + $V_{sw}$ (-2)  & 0.0127   & 0.0015  & 0.335   \\ + $B_Z$ (-2)  & 0.0064   & &  \\ + $B$ (-3)  & 0.0034   & &  \\
\hline
\multicolumn{4}{c}{Parents of AL in 2001 ($k=50$)} \\ Parent ($\tau$ [20 min.]) & $I_{\rm Par \to AL}$ & $I_{\rm SYM-H \to AL}$ & $p$-value  \\ \hline
 No conds.  &  & 0.1388  & $< 10^{-2}$   \\ AL (-1)  & 1.1426   & 0.0094  & $< 10^{-2}$   \\ + $B_Z$ (-1)  & 0.0901   & 0.0059  & $< 10^{-2}$   \\ + AL (-3)  & 0.0159   & 0.0056  & $< 10^{-2}$   \\ + $V_{sw}$ (-1)  & 0.0188   & 0.0034  & 0.035   \\ + $B$ (-3)  & 0.0057   & 0.0030  & 0.070   \\ + $B_Z$ (-2)  & 0.0077   & &  \\ + $P_{dyn}$ (-1)  & 0.0034   & &  \\
\hline
\multicolumn{4}{c}{Parents of AL in 2008 ($k=50$)} \\ Parent ($\tau$ [20 min.]) & $I_{\rm Par \to AL}$ & $I_{\rm SYM-H \to AL}$ & $p$-value  \\ \hline
 No conds.  &  & 0.1651  & $< 10^{-2}$   \\ AL (-1)  & 1.0954   & 0.0088  & $< 10^{-2}$   \\ + $B_Z$ (-1)  & 0.0752   & 0.0078  & $< 10^{-2}$   \\ + $V_{sw}$ (-4)  & 0.0372   & 0.0015  & 0.315   \\ + $P_{dyn}$ (-6)  & 0.0082   & &  \\ + $B_Z$ (-3)  & 0.0035   & &  \\
 \hline
\end{tabular}
\caption{Steps of the causal discovery algorithm (see Methods) for AL and SYM-H for the years 2001 and 2008. The first column lists the iteratively selected conditions. In each step the variable and lag with the highest CMI value (second column) in the preceding step is chosen. The third column gives the CMI value for the substorm -- storm link using the conditions up to this step and the last column its $p$-value. The $p$-values are computed from a block-shuffle ensemble of 200 surrogates (see Methods). The storm -- substorm link vanishes in both years with a $p$-value larger than 0.05 after few solar drivers have been taken into account. The full list of parents in the first column is then used in a second step to estimate ITY (see lag functions in Fig.~\ref{fig:multi_causal} and graphs in Fig.~\ref{fig:graph}). The nearest-neighbor parameter was set to $k=50$, the analysis with very similar results for $k=100$ and another substorm index (AE) is shown in Supplementary Figs.~\ref{fig:multi_causal_SM},\ref{fig:graph_SM} and Tabs.~\ref{tab:significance_analysis_SM_AE_50},\ref{tab:significance_analysis_SM_AE_100}. Here all CMI values are measured in \emph{nats} and are not rescaled to the partial correlation scale as in the other figures.}
\label{tab:significance_analysis}
\end{table}

\begin{figure*}
\centering
\includegraphics[width=\linewidth]{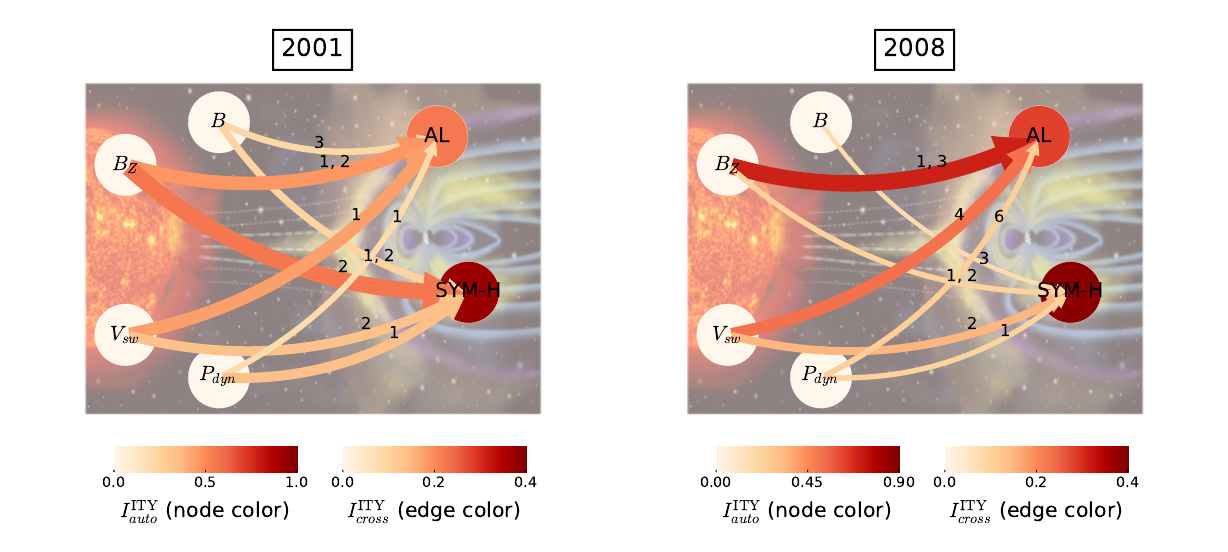}
\caption{Graphs based on significant ITY values at the 95\% level in Fig.~\ref{fig:multi_causal} for 2001 (left) and 2008 (right). Edges correspond to directional lagged links, and the labels indicate their lags. If more than one lag is significant, they are listed in the order of their strength. The edge color and width indicate the value at the lag with the largest ITY. The node color depicts the strength of the lag-1 auto-dependency for AL and SYM-H.
Note that the weak ITY value in $B_Z \to SYM{-}H$ is likely due to $B_Z$ occurring with two neighboring lags in the parents of $SYM{-}H$, which reduces the information transfer of either of them. In Supplementary Fig.~\ref{fig:graph_SM} we show the robustness of these results using a different CMI estimation parameter and another substorm index.}
\label{fig:graph}
\end{figure*}

We begin by investigating bivariate MI and bivTE lag functions of all considered solar variables with AL and SYM-H, including the interaction between these two (gray and black markers in Fig.~\ref{fig:multi_causal}). We restrict the maximum time delay to $\tau_{\max}=6\times20$ min. 
MI lag functions show large values for all possible driver variables in both years. Consider panels $B_Z$ $\to$ AL in both years: Here the peak of the MI lag function is shifted compared to bivTE. Such an effect can arise from strong autocorrelations as studied in Ref.~\cite{Runge2014a}. The bivariate TE has sharper peaks with $B_Z$ clearly being the strongest driver of both AL and SYM-H, and all other drivers are comparably weak (except for auto-dependencies). The reason for this behavior is that some MI values are `inflated', again, due to strong autocorrelations \cite{Runge2012b}, especially $V_{sw}$ is strongly auto-dependent. This makes MI values and the peak of MI lag functions hard to interpret.

The interactions AL $\to$ SYM-H and SYM-H $\to$ AL have been studied in Ref.~\cite{Michelis2011} where a relationship from substorms towards storms was found with a binning estimator of bivTE. Here we reproduce these results with a nearest-neighbor estimator \cite{Kraskov2004a,FrenzelPompe2007}. The other direction, from storms to substorms, is not very significant here. Note that values at lag $\tau=0$ min cannot be interpreted in a directional sense in our analysis.

% \section*{Multivariate causal relationships}

Next, we use a causal discovery algorithm\cite{Runge2012prl} to estimate the preliminary parents, i.e., potential causal drivers, for the multivariate ITY. Table~\ref{tab:significance_analysis} shows iteration steps with the selected parents and the conditional mutual information (CMI) values and significance of the AL $\to$ SYM-H and SYM-H $\to$ AL links in each step. The algorithm described in Ref.~\cite{Runge2012prl} was run with maximum lag $\tau_{\max}=6\times20$ min as before, and we assess significance at the 95\% level (for further details see Methods).
The obtained parents of SYM-H for 2001 and 2008 are very similar except for some differences in the lags. The AL $\to$ SYM-H link becomes non-significant using the condition set ($SYM-H(t-1)$, $B_{Z}(t-2)$, $P_{dyn}(t-1)$, $V_{sw}(t-2)$) in 2001 and with only a different lag in $B_{Z}$ for 2008. This implies that these solar drivers can explain the spurious link AL $\to$ SYM-H at a lag of 20 min. Note that this set is only a sufficient explanatory set and other drivers might also induce this spurious association.
Also the much weaker link SYM-H $\to$ AL becomes non-significant after including few solar drivers ($B_{Z}$, $V_{sw}$ at different lags in both years, in 2001 also $B$). 

The ITY estimates with these  parents are shown in Fig.~\ref{fig:multi_causal} (blue markers). ITY now accounts for autocorrelation in the driven variable (like bivTE), but additionally for the influence of the other parents as common drivers or indirect mediators \cite{Runge2015}. Now the ITY lag functions are peaked only at a few selected lags: the strongest causal influence to both AL and SYM-H comes from their respective past lags, but the second strongest driver is $B_Z$ which drives AL at lag 20 min and SYM-H at lag 40 min in both years (in 2008 a lag of 20 min is also significant for SYM-H). Note that $B_Z$ occurs twice at different lags in the ITY conditions of SYM-H in 2008 and of AL in 2001 and 2008. Their two ITY values are, therefore, smaller because the information is shared among both lags.
In Fig.~\ref{fig:graph} we visualize the significant drivers of AL and SYM-H in a process graph as in Supplementary Fig.~\ref{fig:tsg}.

%%
%% Discussion
%%
\section*{Discussion}

DeMichelis \emph{et al.} \cite{Michelis2011} investigated the transfer of information between substorms and storms by means of a bivariate transfer entropy analysis of AL and SYM-H time series from 1981 (near solar maximum). They found a significant information flow from substorms to storms attaining its maximum with a typical time delay of about 1 h and suggested that the direction of information flow between substorms and storms depends on the global magnetospheric activity level. 
Our analysis goes beyond the study of Ref.~\cite{Michelis2011} by utilizing a directional, multivariate information-theoretic causality measure that simultaneously takes into account solar wind variables and geomagnetic indices data, allowing for more powerful statistical tests on the absence or potential presence of a causal relationship between substorms and storms.

Our major results can be summarized as follows: 
The main drivers of substorms as measured by AL are $B_Z$ and $V_{SW}$. These also drive storms as measured by SYM-H. This is consistent with the fact that the energy transfer from the solar wind to the magnetosphere is proportional to $B_Z$ and $V_{SW}$. 
$P_{dyn}$ and especially $B$ are less relevant for both storms and substorms.
Regarding time lags, the AL index first responds to $B_Z$ (lag 20--40 min), while the lags with the weaker other drivers are varying between the different years. The SYM-H index also first responds to $B_Z$ (lag 20--40 min) and $P_{dyn}$ (20 min), then to $V_{SW}$ (40 min) and rather weakly with non-robust lags to $B$ (see Supplementary Figs.~\ref{fig:multi_causal_SM},\ref{fig:graph_SM}). Since we find these drivers to be very similar during the near solar maximum (2001) and minimum (2008) years for both storms and substorms, we conclude that these directed information transfers are rather independent of the activity level and constitute robust interrelationships between solar wind parameters and dynamic processes in the magnetosphere. 

Most importantly, our iterative causal discovery algorithm analysis suggests that mainly $B_Z$, and to a lesser degree $V_{SW}$ and $P_{dyn}$ are sufficient to explain the previously found link AL $\to$ SYM-H \cite{Michelis2011}. These results are also verified by applying  the same tools to an AE - SYM-H analysis and for other estimation parameters (see Supplementary Figs.~\ref{fig:multi_causal_SM},\ref{fig:graph_SM} and Tabs.~\ref{tab:significance_analysis_SM_AE_50},\ref{tab:significance_analysis_SM_AE_100}).
Thus, we find that there is no direct or indirect transfer of information AL $\to$ SYM-H or SYM-H $\to$ AL.
Note that the presence of significant links in our analysis can be called causal only with respect to the included set of variables. Non-observed variables can still be the cause of a link here. What is more certain is non-causality: if there exists no statistical evidence for a dependency between two variables, a physical mechanism between the two is unlikely.

Although some studies \cite{Daglis2000,Gkioulidou2014}, based on in-situ observations, have shown that the contribution of ion injections to the ring current energy gain is substantial, our results refute the role of substorms in the enhancement of the storm-time ring current through accumulative ion injections during consecutive substorms. A possible reason for the absence of information transfer from AL $\to$ SYM-H might be the possibility that not all ion injections to the storm-time ring current are reflected in the AL variations. A recent study \cite{Gkioulidou2015} showed that small-scale injections are not captured by AL. Another study \cite{Gkioulidou2016} showed that low- and high-energy protons vary in quite different ways on storm-time timescales and accordingly suggested that the relation between ion injections and ring current growth may be more complicated than previously perceived.

Nevertheless, the non-significance of direct or indirect dependencies between the commonly used AL and SYM-H indices leads us to the conclusion that there exists no physical mechanism by which perturbations in substorms are transported to storms or vice versa. However, their non-significance may have several other reasons: 
Firstly, the information measure might not capture the dependency. We should note that our information-theoretic approach allows to take into account almost any type of nonlinear relationship, both in excluding it as a common driver, and also in detecting it. The price for this ``universality'' is lower statistical power, which means that weaker dependencies might not be detected. However, here we have a very large sample size leading us to the conclusion that if there is a dependency, it must be very weak. Also, our results are robust when using other estimation parameters (see Supplementary Figs.~\ref{fig:multi_causal_SM},\ref{fig:graph_SM} and Tabs.~\ref{tab:significance_analysis_SM_AE_50},\ref{tab:significance_analysis_SM_AE_100}).
Secondly, we analyzed the whole years 2001 and 2008 as two extremes of solar activity to obtain a sufficiently large sample size. Possibly, a causal relationship is present only during shorter periods, which would be difficult to assess given too short sample sizes and the length of characteristic time scales of the processes. 
Thirdly, the physical mechanism might be present mostly during the missing values excluded in the analysis. If satellite failures are indeed strongly related with the hypothesized mechanism, this would imply a non-avoidable selection bias in our analysis. 
Lastly, the indices might not be good proxies of substorms or storms or be contaminated by measurement noise. Here we tested two kinds of indices for substorms (AL and AE) and got robust results.
In light of these qualifications, we conclude that a direct or indirect physical mechanism by which substorms drive storms or vice versa is unlikely.

\section*{Conclusions}

Our analysis demonstrates the great potential of combining a causal discovery algorithm with a multivariate and lag-specific extension of transfer entropy for tackling contemporary research questions in magnetospheric physics, such as the storm-substorm relationship, which is one of the most controversial topics of magnetospheric dynamics and solar-terrestrial coupling. Further analyses using a causal pathway-analysis \cite{Runge2014c,Runge2015} can shed  light on the interaction mechanism among the solar drivers and the magnetosphere. The obtained causal drivers, on the other hand, can also be relevant for optimal prediction schemes \cite{Runge2014b}.
We expect that our results will contribute to a better understanding of the dynamic processes related to the coupled solar wind - magnetosphere - ionosphere system by fostering a paradigm shift in our perception of the storm-substorm relationship. They may also have a direct impact on magnetosphere modeling and, consequently, space weather forecasting efforts.

\section*{Methods}

\subsection*{Data}
The present study is focused on two years (2001 and 2008) of data on solar wind parameters and geomagnetic activity indices (downloaded from http://omniweb.gsfc.nasa.gov/form/omni\_min.html). As possible solar driving factors we include only those quantities with at least a measurable statistical dependency (mutual information) with either AL or SYM-H. These are $B$, $B_Z$, $V_{SW}$, and $P_{dyn}$. The original one-minutely data were aggregated to a 20-minute time resolution by averaging over non-overlapping 20-minute blocks. This resolution was selected based on iterative tests to obtain a compromise between resolving time lags and still keeping the computational load and multiple testing problems low. Additionally, DeMichelis \emph{et al.} \cite{Michelis2011} found, on average, a net information flow from AL to SYM-H attaining its maximum at a typical time delay of about 1h which is well resolved with our chosen time resolution.

As solar wind time series inevitably contain missing values due to satellite failures, in the aggregation we masked samples for 20 min periods with more than 50\% missing values (Supplementary Fig.~\ref{fig:data}). We also accounted for masked samples in the lagged analyses (up to $\tau_{\max}=6\times20$ min) to avoid a selection bias. For the solar maximum year of 2001 this leads to 18,384 non-masked 20-min samples and for the solar minimum year of 2008 to 17,893 samples instead of about 26,000 samples for the whole year. No further pre-processing was applied. Supplementary Fig.~\ref{fig:data} shows the corresponding time series.

\subsection*{Information-theoretic causality analysis}
This study aims to shed light on the possible existence of a driver-response relationship between storms and substorms, which is a reflection of the dynamic processes within the coupled solar wind–magnetosphere–ionosphere system. Because there has been accumulating evidence that the involved interrelations are of a nonlinear nature \cite{Bargatze1985,Tsurutani1990,Consolini1996,Klimas1996,Vassiliadis1996,Hnat2002,Balasis2009} and very long data series are available, we employ a non-parametric (model-free) approach here. Information theory provides a genuine framework for the model-free study of couplings among time series.
Here we invoke three information-theoretic measures with increasing power to detect spurious dependencies due to autocorrelation, common drivers or indirect relationships.

The first and simplest association measure applying information theory to time series is the lagged (cross-)mutual information \cite{Cover2006} given by 
\begin{align} \label{eq:lagged_mi}
I^{\rm MI}_{XY}(\tau) = I(X_{t-\tau};Y_t) = H(Y_t) - H(Y_t~|~X_{t-\tau})\,,
\end{align}
using Shannon entropies $H={-}\int p(x) \ln p(x) dx$ with the natural logarithm. All information measures are here studied in the corresponding units of \emph{nats}. For $\tau>0$, MI measures the information in the past of $X$ that is contained in the present of $Y$. 
The weaknesses of MI as a measure of information transfer have been discussed early on, most notably by Schreiber \cite{Schreiber2000b}. A first step to arrive at a directional notion of information transfer is to exclude information from the past of $Y$. Implementing this idea, Schreiber introduced the \emph{transfer entropy} (TE) \cite{Schreiber2000b} between two variables, which is the information-theoretic analogue of Granger causality and can be defined in a lag-specific variant as
\begin{align} \label{eq:def_te}
I_{X \to Y}^{\rm bivTE}(\tau) &= I(X_{t-\tau};Y_t~|~Y_{t-1})
\end{align}
based on the \emph{conditional mutual information}.

Notably, bivTE can yield spurious results if more than two processes are interacting: For the interaction example in Supplementary Fig.~\ref{fig:tsg}(b) both the MI $I(X_{t-1};Y_t)$ and the TE $I(X_{t-1};Y_t|Y_{t-1})$ are larger than zero due to the common driver $Z$ even though no direct or indirect physical mechanism exists by which $X$ drives $Y$ or vice versa. The detailed time-resolved graph in Supplementary Fig.~\ref{fig:tsg}(a) shows that, $X_{t-1}$ and $Y_{t}$ are \emph{not} independent given only the past of $Y$ or, on the other hand, only the common driver $Z_{t-2}$ as a condition. Rather, both conditions need to be included to exclude all causal paths connecting the two and unveil the spurious dependency (see Ref.~\cite{Runge2015} for a definition of causal paths). 

These subtle interactions can be captured with the concept of a \emph{time series graph} \cite{Dahlhaus2000,Eichler2011} as shown in Supplementary Fig.~\ref{fig:tsg}(a), originating from the theory of graphical models.
As further defined in Ref.~\cite{Runge2015}, each node in a time series graph represents a subprocess at a certain time. Past nodes at $t'<t$ have a link towards a subprocess at time $t$ if and only if they are not independent conditionally on the past of the whole process, which implies a lag-specific Granger causality with respect to the measured process. In this graph the parents $\mathcal{P}_{\bullet}$ of a variable are given by all nodes with an arrow towards it (blue boxes in Supplementary Fig.~\ref{fig:tsg}(a)). 

While these parents could be estimated by testing the CMI between each $X_{t-\tau}$ and $Y_t$ conditional on all other lagged variables, this approach, similar to multivariate or conditional TE, does not work well due to its high dimensionality \cite{Runge2012prl} leading to weak statistical power and many false positives.

In Ref.~\cite{Runge2012prl} an efficient algorithm has been introduced for the estimation of the parents of a variable $Y$. The original idea is to successively test for conditional independence between $Y_t$ and each possible past driver (including the past of $Y$) conditioned on iteratively more conditions and testing all combinations of conditions. Thereby, the dimension stays as low as possible in every iteration step which helps to alleviate high dimensionality in estimating CMIs. However this comes at a considerable computational cost which is extremely high in our case due to the large sample size of order $\mathcal{O}(10^5)$. Therefore, here we add only the most relevant condition with the highest CMI in the previous step (setting $n_i=1$ in the causal discovery algorithm). The drawback is that the algorithm now might still contain non-causal spurious drivers \cite{Runge2014b,Runge2015}, which necessitates a second step: the preliminary set of parents is used to estimate the \emph{information transfer to Y} (ITY) \cite{Runge2012b} for all lagged variables $X_{t-\tau}$ (including the parents)
\begin{align} \label{eq:def_ity}	
I_{X \to Y}^{\rm ITY}(\tau) &= I(X_{t-\tau};Y_t~|~\mathcal{P}_{Y_t}) \,,
\end{align}
which will be zero if and only if $X_{t-\tau}$ and $Y_t$ are independent \emph{conditionally on $\mathcal{P}_{Y_t}$}.

Here we use an advanced nearest-neighbor estimator \cite{Kraskov2004a,FrenzelPompe2007} of CMI that is most suitable for variables with a continuous range of values. This estimator has as a parameter the number of nearest-neighbors $k$ which determines the size of hyper-cubes around each (high-dimensional) sample point and, therefore, can be viewed as a density smoothing parameter (although it is data-adaptive unlike fixed-bandwidth estimators). For large $k$, the underlying dependencies are strongly smoothed. We tested different values of $k$ to verify the robustness of our results. Larger $k$ have larger bias and are more computationally expensive, but have smaller variance, which is more important for significance testing. Note that the estimated CMI values can be slightly negative while CMI is a non-negative quantity. In Figs.~\ref{fig:multi_causal},\ref{fig:graph},\ref{fig:multi_causal_SM},\ref{fig:graph_SM} the (C)MI values have been rescaled to the (partial) correlation scale via $I \to\sqrt{1-e^{-2I}}\in [0,1]$ \cite{Cover2006}. In the tables, on the other hand, the CMI values are given in \emph{nats}.

Unfortunately, no analytical results exist on the finite-sample distribution of this estimator under the null hypothesis of conditional independence. For significance testing, either a fixed threshold or shuffle surrogates are, therefore, the only choice here. Surrogate tests are especially helpful for proper significance tests because they adapt to the bias for higher-dimensional CMIs.
In Ref.~\cite{Runge2012prl} a shuffle test has been used, but for strongly autocorrelated time series, as in the present case, this test is too weak. Therefore, we use a block-shuffle surrogate test here following Ref.~\cite{Peifer2005} and Ref.~\cite{Mader2013}. An ensemble of $M$ values of $I(X^*_{t-\tau}; Y_t\,|\,Z)$ is generated where $X^*_{t-\tau}$ is a block-shuffled sample of $X_{t-\tau}$, i.e., with blocks of the original time series permuted. As an optimal block-length we use the approach described in Ref.~\cite{Peifer2005} and Ref.~\cite{Mader2013} for non-overlapping blocks. The optimal block-length (Eq.~(6) in Ref.~\cite{Mader2013}) involves the decay rate of the envelope of the autocorrelation function $\gamma(\tau)$. The latter is estimated up to a maximum delay of 5\% of the (non-masked) samples and the envelope was estimated using the Hilbert transform. Then a function $C \phi^{\tau}$ is fitted to the envelope with constant $C$ to obtain the decay rate $\phi$.
Finally, the CMI values are sorted and a $p$-value is obtained as the fraction of surrogates with CMI greater or equal than the estimated CMI value. We use an ensemble of $200$ surrogates. Confidence intervals (errorbars in figures) were estimated using bootstrap resampling involving only estimated nearest-neighbor statistics with $200$ samples.

The time series graph in Supplementary Fig.~\ref{fig:tsg}(a) closely depicts the potentially causal interactions found in this study with $X$ standing for AL, $Z$ for $B_Z$, and $Y$ for SYM-H. The causal information flow along directed links in this graph can well explain the spurious `significant' MI and bivTE values found in Ref.~\cite{Michelis2011}. Also the small bivTE value in the other direction, SYM-H $\to$ AL, can be explained by these solar drivers.

%
%Topical subheadings are allowed. Authors must ensure that their %Methods section includes adequate experimental and characterization %data necessary for others in the field to reproduce their work.

% \bibliography{sample}

% \noindent LaTeX formats citations and references automatically using the bibliography records in your .bib file, which you can edit via the project menu. Use the cite command for an inline citation, e.g.  \cite{Figueredo:2009dg}.

\subsection*{Software availability}
Software is available online under \verb|https://github.com/jakobrunge/tigramite|.

\bibliography{Runge.bib}

\section*{Acknowledgements}
J.R. received funding by the James S. McDonnell Foundation. R.V.D. acknowledges financial support via the BMBF-funded Young Investigators Group ``Complex Systems Approaches to Understanding Causes and Consequences of Past, Present and Future Climate Change'' (CoSy-CC$^2$, grant no. 01LN1306A). This work has been financially supported by the joint Greek–German IKYDA 2013 project ``Transdisciplinary assessment of dynamical complexity in magnetosphere and climate: A unified description of the nonlinear dynamics across extreme events'' funded by IKY and DAAD.
The authors thank Ciaron Linstead for help with high-performance computing and gratefully acknowledge the European Regional Development Fund (ERDF), the German Federal Ministry of Education and Research (BMBF) and the Land Brandenburg for supporting this project by providing resources on the high performance computer system at the Potsdam Institute for Climate Impact Research.

\section*{Author contributions}
G.B., J.R., R.V.D., and I.A.D. designed the study, C.P. prepared the data, J.R. analyzed the data. All authors discussed the results and contributed to editing the manuscript.

\section*{Additional information}
Supplementary information is available in the Supplementary Material of the paper.

%TODO 
\section*{Competing financial interests} 
The authors declare no competing interests. 

% The corresponding author is responsible for submitting a \href{http://www.nature.com/srep/policies/index.html#competing}{competing financial interests statement} on behalf of all authors of the paper. This statement must be included in the submitted article file.

%%
%%  SUPPLEMENTARY INFORMATION
%%
\clearpage
% \onecolumn
\renewcommand\theequation{S\arabic{equation}}
\setcounter{equation}{0}
\renewcommand\thefigure{S\arabic{figure}}    
\setcounter{figure}{0}   
\renewcommand\thesection{S\arabic{section}}    
\setcounter{section}{0}  
\renewcommand\thetable{S\arabic{table}}    
\setcounter{table}{0}  

\section*{{\Huge Supplementary Material}
\\
\\
for ``Common solar wind drivers behind magnetic storm--magnetospheric substorm dependency''
\\
\\
by Jakob Runge, Georgios Balasis, Ioannis A. Daglis, Constantinos Papadimitriou,
and Reik V. Donner
}

\clearpage
\begin{figure*}
\centering
\includegraphics[width=\linewidth]{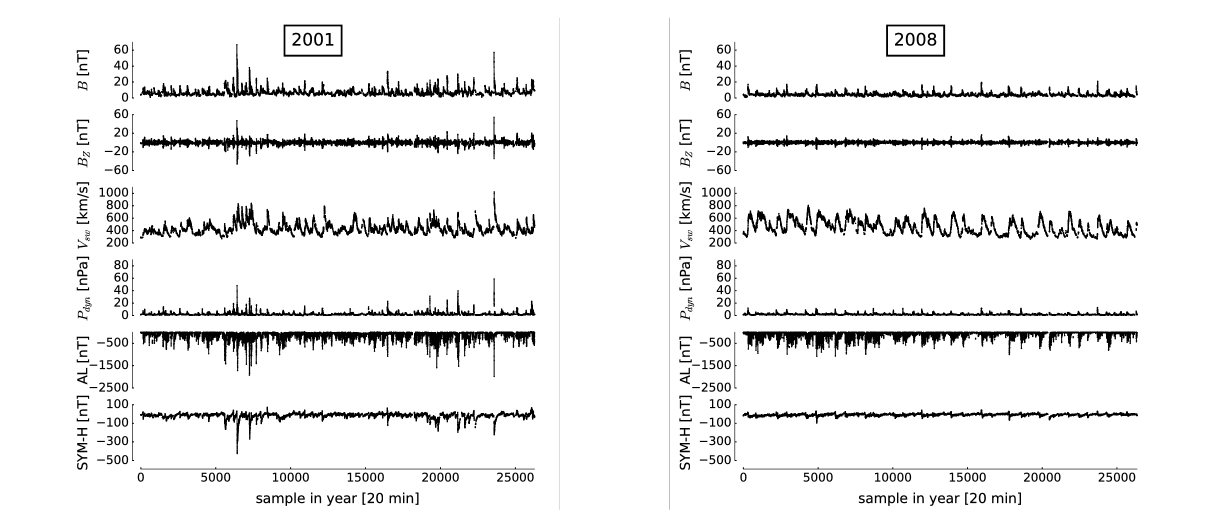}
\caption{Time series for 2001 (left) and 2008 (right). Time points with missing values in any of the variables are excluded from the analysis, taking lags into account. Clearly, there is much less variability in all variables during the solar minimum year 2008.}
\label{fig:data}
\end{figure*}

\clearpage

\begin{figure*}
\centering
\includegraphics[width=.5\linewidth]{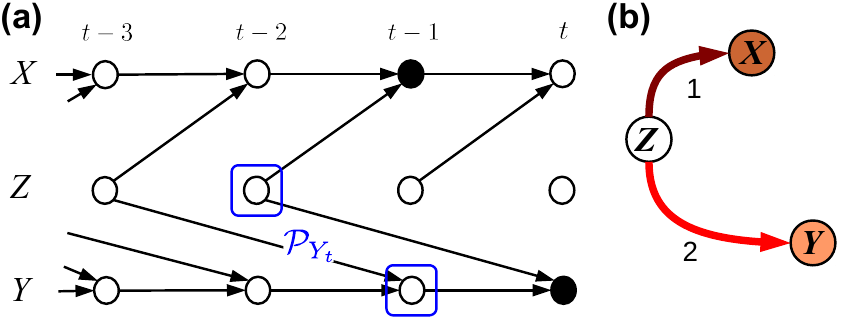}
\caption{Example of causal interactions in a three-variable process. 
(a) Time series graph which encodes the spatio-temporal dependencies. The set of \textit{parents} $\mathcal{P}_{Y_t}$ (blue boxes) separates $Y_t$ from the past of the whole process $\mathbf{X}_t^- {\setminus} \mathcal{P}_{Y_t}$, which implies conditional independence (Markov property) and is used in the algorithm \cite{Runge2012prl} to estimate the graph.
(b) Process graph, which aggregates the information in the time series graph for better visualization (labels denote the lags, link and node colors denote the cross- and auto-coupling strength).}
\label{fig:tsg}
\end{figure*}

\clearpage
\begin{figure*}
\centering
\includegraphics[width=.86\linewidth]{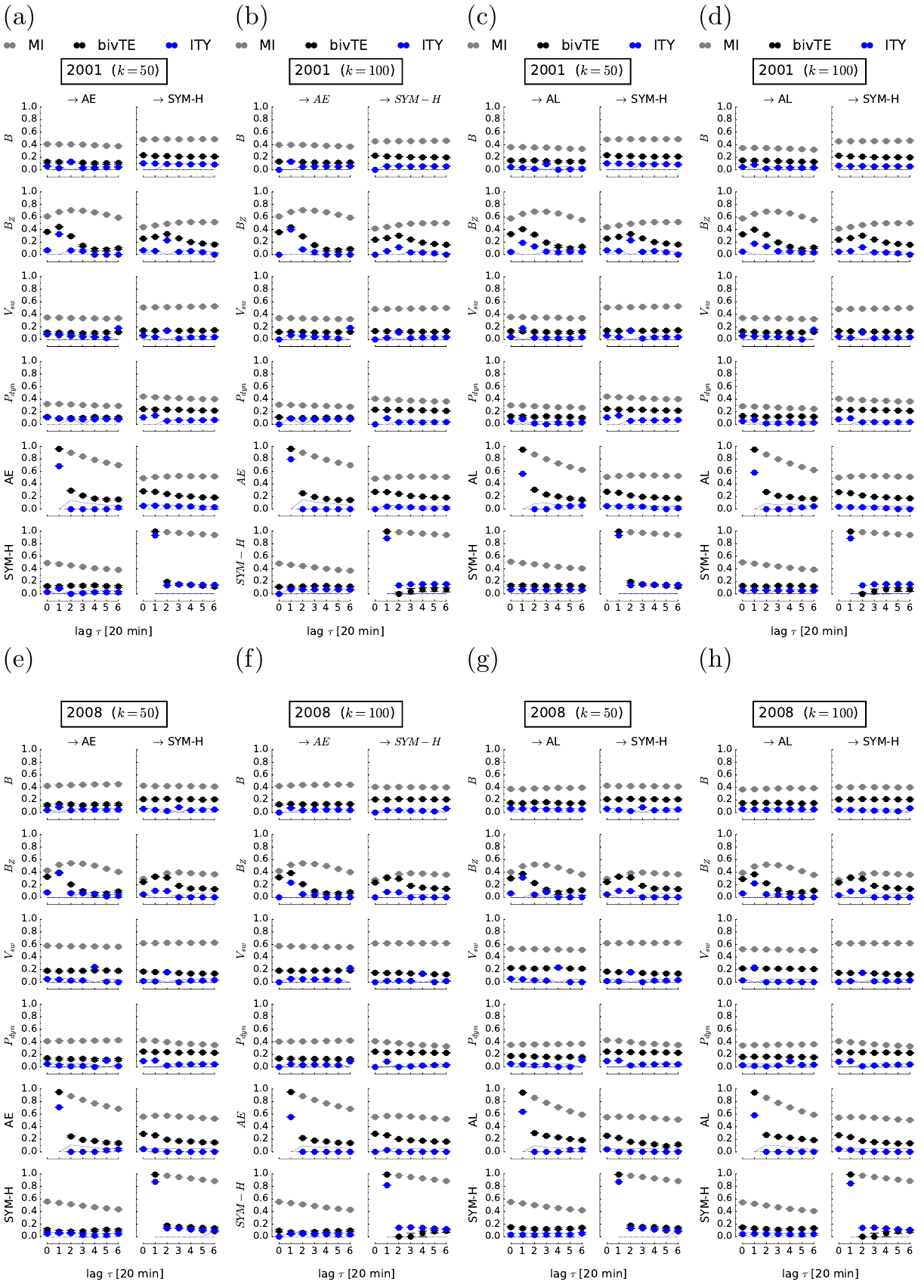}
\caption{Lag functions as in Fig.~\ref{fig:multi_causal}, but for additional parameters: year 2001 (a-d) and 2008 (e-h); nearest-neighbor CMI estimation parameter $k=50$ (a,c,e,g) and $k=100$ (b,d,f,h) resulting in a stronger smoothing of the densities; substorm index AE (a,b,e,f) and AL (c,d,g,h).}
\label{fig:multi_causal_SM}
\end{figure*}

\clearpage
\begin{figure*}
\centering
\includegraphics[width=.65\linewidth]{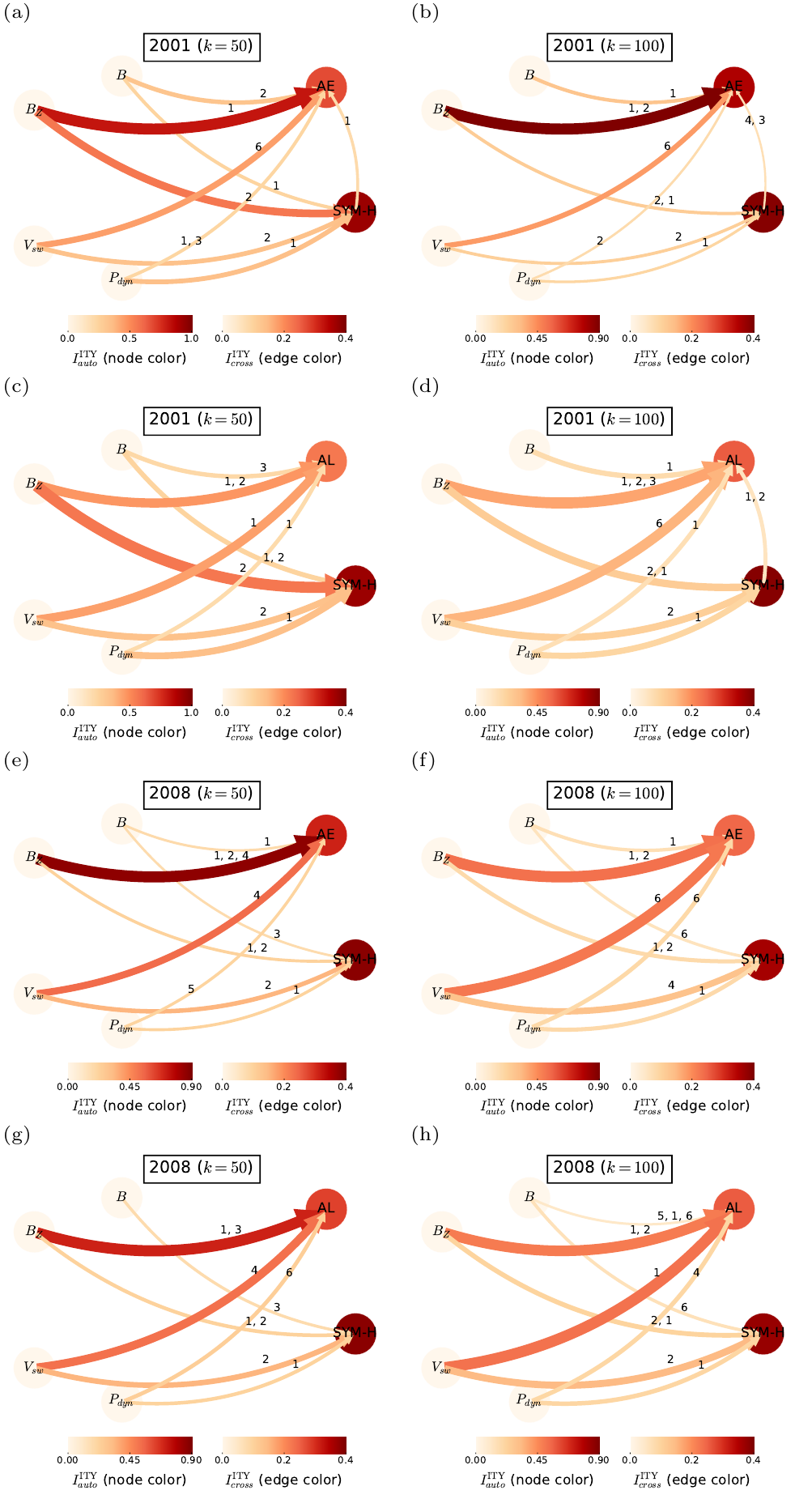}
\caption{As in Fig.~\ref{fig:graph}, but for additional parameters: year 2001 (a-d) and 2008 (e-h); nearest-neighbor CMI estimation parameter $k=50$ (a,c,e,g) and $k=100$ (b,d,f,h) resulting in a stronger smoothing of the densities; substorm index AE (a,b,e,f) and AL (c,d,g,h).}
\label{fig:graph_SM}
\end{figure*}

\clearpage

\begin{table}[t!]
\centering
\begin{tabular}{l|c|c|c}  
\hline
\hline
\multicolumn{4}{c}{Parents of SYM-H in 2001 ($k=100$)} \\ Parent ($\tau$ [20 min.]) & $I_{\rm Par \to SYM-H}$ [nats] & $I_{\rm AL \to SYM-H}$ [nats] & $p$-value  \\ \hline
 No conds.  &  & 0.1569  & $< 10^{-2}$   \\ SYM-H (-1)  & 2.2162   & 0.0336  & $< 10^{-2}$   \\ + $B_Z$ (-2)  & 0.0479   & 0.0069  & $< 10^{-2}$   \\ + $P_{dyn}$ (-1)  & 0.0126   & 0.0018  & $< 10^{-2}$   \\ + $V_{sw}$ (-2)  & 0.0077   & 0.0010  & 0.145   \\ + $B_Z$ (-1)  & 0.0016   & &  \\
\hline
\multicolumn{4}{c}{Parents of SYM-H in 2008 ($k=100$)} \\ Parent ($\tau$ [20 min.]) & $I_{\rm Par \to SYM-H}$ [nats] & $I_{\rm AL \to SYM-H}$ [nats] & $p$-value  \\ \hline
 No conds.  &  & 0.1826  & $< 10^{-2}$   \\ SYM-H (-1)  & 1.9224   & 0.0280  & $< 10^{-2}$   \\ + $B_Z$ (-1)  & 0.0502   & 0.0113  & $< 10^{-2}$   \\ + $P_{dyn}$ (-1)  & 0.0132   & 0.0041  & $< 10^{-2}$   \\ + $V_{sw}$ (-2)  & 0.0110   & 0.0019  & $< 10^{-2}$   \\ + $B_Z$ (-2)  & 0.0055   & 0.0001  & 0.880   \\ + $B$ (-6)  & 0.0023   & &  \\
\hline
\multicolumn{4}{c}{Parents of AL in 2001 ($k=100$)} \\ Parent ($\tau$ [20 min.]) & $I_{\rm Par \to AL}$ [nats] & $I_{\rm SYM-H \to AL}$ [nats] & $p$-value  \\ \hline
 No conds.  &  & 0.1330  & $< 10^{-2}$   \\ AL (-1)  & 1.1489   & 0.0079  & $< 10^{-2}$   \\ + $B_Z$ (-1)  & 0.0865   & 0.0052  & $< 10^{-2}$   \\ + AL (-4)  & 0.0125   & 0.0051  & $< 10^{-2}$   \\ + $V_{sw}$ (-6)  & 0.0136   & 0.0025  & 0.035   \\ + $B_Z$ (-2)  & 0.0078   & 0.0018  & 0.060   \\ + $B$ (-1)  & 0.0045   & &  \\ + $P_{dyn}$ (-1)  & 0.0026   & &  \\
\hline
\multicolumn{4}{c}{Parents of AL in 2008 ($k=100$)} \\ Parent ($\tau$ [20 min.]) & $I_{\rm Par \to AL}$ [nats] & $I_{\rm SYM-H \to AL}$ [nats] & $p$-value  \\ \hline
 No conds.  &  & 0.1599  & $< 10^{-2}$   \\ AL (-1)  & 1.1013   & 0.0087  & $< 10^{-2}$   \\ + $B_Z$ (-1)  & 0.0723   & 0.0073  & $< 10^{-2}$   \\ + $V_{sw}$ (-1)  & 0.0372   & 0.0005  & 0.605   \\ + $P_{dyn}$ (-4)  & 0.0058   & &  \\ + $B_Z$ (-3)  & 0.0038   & &  \\ + $B_Z$ (-2)  & 0.0013   & &  \\
  \hline
\end{tabular}
\caption{As in Tab.~\ref{tab:significance_analysis}, but for a nearest-neighbor CMI estimation parameter $k=100$, i.e., with a stronger smoothing of the densities.}
\label{tab:significance_analysis_SM_AL}
\end{table}

\clearpage
\begin{table}[t!]
\centering
\begin{tabular}{l|c|c|c}  
\hline
\hline
\multicolumn{4}{c}{Parents of SYM-H in 2001 ($k=50$)} \\ Parent ($\tau$ [20 min.]) & $I_{\rm Par \to SYM-H}$ & $I_{\rm AE \to SYM-H}$ & $p$-value  \\ \hline
 No conds.  &  & 0.1527  & $< 10^{-2}$   \\ SYM-H (-1)  & 2.2092   & 0.0395  & $< 10^{-2}$   \\ + $B_Z$ (-2)  & 0.0571   & 0.0100  & $< 10^{-2}$   \\ + $P_{dyn}$ (-1)  & 0.0164   & 0.0032  & $< 10^{-2}$   \\ + $V_{sw}$ (-2)  & 0.0094   & 0.0016  & 0.405   \\
\hline
\multicolumn{4}{c}{Parents of SYM-H in 2008 ($k=50$)} \\ Parent ($\tau$ [20 min.]) & $I_{\rm Par \to SYM-H}$ & $I_{\rm AE \to SYM-H}$ & $p$-value  \\ \hline
 No conds.  &  & 0.2003  & $< 10^{-2}$   \\ SYM-H (-1)  & 1.9099   & 0.0355  & $< 10^{-2}$   \\ + $B_Z$ (-1)  & 0.0569   & 0.0159  & $< 10^{-2}$   \\ + $P_{dyn}$ (-1)  & 0.0162   & 0.0059  & $< 10^{-2}$   \\ + $V_{sw}$ (-2)  & 0.0127   & 0.0027  & 0.075   \\ + $B_Z$ (-2)  & 0.0064   & &  \\ + $B$ (-3)  & 0.0034   & &  \\
\hline
\multicolumn{4}{c}{Parents of AE in 2001 ($k=50$)} \\ Parent ($\tau$ [20 min.]) & $I_{\rm Par \to AE}$ & $I_{\rm SYM-H \to AE}$ & $p$-value  \\ \hline
 No conds.  &  & 0.1281  & $< 10^{-2}$   \\ AE (-1)  & 1.2574   & 0.0074  & $< 10^{-2}$   \\ + $B_Z$ (-1)  & 0.1088   & 0.0074  & $< 10^{-2}$   \\ + $B_Z$ (-3)  & 0.0093   & 0.0050  & $< 10^{-2}$   \\ + $B$ (-2)  & 0.0085   & 0.0045  & 0.015   \\ + $V_{sw}$ (-6)  & 0.0177   & 0.0036  & 0.030   \\ + SYM-H (-1)  & 0.0036   & &  \\
\hline
\multicolumn{4}{c}{Parents of AE in 2008 ($k=50$)} \\ Parent ($\tau$ [20 min.]) & $I_{\rm Par \to AE}$ & $I_{\rm SYM-H \to AE}$ & $p$-value  \\ \hline
 No conds.  &  & 0.1706  & $< 10^{-2}$   \\ AE (-1)  & 1.1705   & 0.0032  & $< 10^{-2}$   \\ + $B_Z$ (-1)  & 0.0841   & 0.0068  & $< 10^{-2}$   \\ + $V_{sw}$ (-4)  & 0.0426   & 0.0021  & 0.350   \\ + $B$ (-1)  & 0.0102   & &  \\ + $P_{dyn}$ (-5)  & 0.0050   & &  \\
   \hline
\end{tabular}
\caption{As in Tab.~\ref{tab:significance_analysis_SM_AL}, but for another substorm index (AE) and $k=50$.}
\label{tab:significance_analysis_SM_AE_50}
\end{table}

\clearpage
\begin{table}[t!]
\centering
\begin{tabular}{l|c|c|c}  
\hline
\multicolumn{4}{c}{Parents of SYM-H in 2001 ($k=100$)} \\ Parent ($\tau$ [20 min.]) & $I_{\rm Par \to SYM-H}$ & $I_{\rm AE \to SYM-H}$ & $p$-value  \\ \hline
 No conds.  &  & 0.1469  & $< 10^{-2}$   \\ SYM-H (-1)  & 2.2162   & 0.0382  & $< 10^{-2}$   \\ + $B_Z$ (-2)  & 0.0479   & 0.0079  & $< 10^{-2}$   \\ + $P_{dyn}$ (-1)  & 0.0125   & 0.0024  & $< 10^{-2}$   \\ + $V_{sw}$ (-2)  & 0.0077   & 0.0012  & 0.125   \\ + $B_Z$ (-1)  & 0.0016   & &  \\
\hline
\multicolumn{4}{c}{Parents of SYM-H in 2008 ($k=100$)} \\ Parent ($\tau$ [20 min.]) & $I_{\rm Par \to SYM-H}$ & $I_{\rm AE \to SYM-H}$ & $p$-value  \\ \hline
 No conds.  &  & 0.1973  & $< 10^{-2}$   \\ SYM-H (-1)  & 1.9225   & 0.0361  & $< 10^{-2}$   \\ + $B_Z$ (-1)  & 0.0502   & 0.0134  & $< 10^{-2}$   \\ + AE (-1)  & 0.0134   & 0.0134  & $< 10^{-2}$   \\ + $V_{sw}$ (-4)  & 0.0087   & 0.0134  & $< 10^{-2}$   \\ + $P_{dyn}$ (-1)  & 0.0063   & 0.0134  & $< 10^{-2}$   \\ + $B_Z$ (-2)  & 0.0034   & 0.0134  & $< 10^{-2}$   \\ + $B$ (-6)  & 0.0020   & &  \\
\hline
\multicolumn{4}{c}{Parents of AE in 2001 ($k=100$)} \\ Parent ($\tau$ [20 min.]) & $I_{\rm Par \to AE}$ & $I_{\rm SYM-H \to AE}$ & $p$-value  \\ \hline
 No conds.  &  & 0.1205  & $< 10^{-2}$   \\ AE (-1)  & 1.2673   & 0.0067  & $< 10^{-2}$   \\ + $B_Z$ (-1)  & 0.1056   & 0.0060  & $< 10^{-2}$   \\ + $B$ (-1)  & 0.0102   & 0.0045  & $< 10^{-2}$   \\ + $V_{sw}$ (-6)  & 0.0174   & 0.0026  & 0.080   \\
\hline
\multicolumn{4}{c}{Parents of AE in 2008 ($k=100$)} \\ Parent ($\tau$ [20 min.]) & $I_{\rm Par \to AE}$ & $I_{\rm SYM-H \to AE}$ & $p$-value  \\ \hline
 No conds.  &  & 0.1684  & $< 10^{-2}$   \\ AE (-1)  & 1.1764   & 0.0024  & 0.010   \\ + $B_Z$ (-1)  & 0.0811   & 0.0067  & $< 10^{-2}$   \\ + $V_{sw}$ (-6)  & 0.0404   & 0.0014  & 0.315   \\ + $B$ (-1)  & 0.0089   & &  \\ + $P_{dyn}$ (-6)  & 0.0043   & &  \\ + $B_Z$ (-3)  & 0.0028   & &  \\ + $B_Z$ (-2)  & 0.0013   & &  \\
  \hline
\end{tabular}
\caption{As in Tab.~\ref{tab:significance_analysis_SM_AL}, but for another substorm index (AE)  and $k=100$.}
\label{tab:significance_analysis_SM_AE_100}
\end{table}

\end{document}